# Title: From local collective behavior to global migratory patterns in white storks

**Authors:** Andrea Flack[a,b†*], Máté Nagy[b,c,d†*], Wolfgang Fiedler[a,b], Iain D. Couzin[b,c], Martin Wikelski[a,b]

**Affiliations:**
[a]Department of Migration and Immuno-Ecology, Max Planck Institute for Ornithology, Radolfzell, Germany.
[b]Department of Biology, University of Konstanz, Konstanz, Germany.
[c]Department of Collective Behaviour, Max Planck Institute for Ornithology, Konstanz, Germany.
[d]MTA-ELTE Statistical and Biological Physics Research Group, Hungarian Academy of Sciences, Budapest, Hungary

[†] These authors contributed equally to this work.
[*] Corresponding authors: aflack@orn.mpg.de; mnagy@orn.mpg.de

**Abstract:** Soaring migrants exploit columns of rising air (thermals) to cover large distances with minimal energy. Employing social information while locating thermals may be beneficial, but examining collective movements in wild migrants has been a major challenge. We investigated the group movements of a flock of 27 naturally migrating, juvenile white storks using high-resolution GPS and accelerometers. Analyzing individual and group movements on multiple scales reveals that birds ahead navigate to and explore thermals, while followers benefit from their movements. Despite this benefit, followers often leave thermals earlier and at lower height, and consequently must flap considerably more. Followers also migrated less far annually than did leaders. We provide insights into the interactions between freely-flying social migrants and the costs and benefits of collective movement in natural populations.

**One Sentence Summary:** Differences in leader-follower behavior and flight effort during fine-scale navigation within, and between thermals, relate to major differences in storks' global migration**.**

**Main Text:** In many animal taxa, migrations are performed by large social groups (*1*) providing various benefits to individual group members (*2*–*6*). In heterogeneous groups, coordinated movements frequently generate leader-follower patterns (*7*–*10*), and individuals may adopt these behavioral strategies, presumably without knowing their own or their group members' roles (*11*). Given inter-individual variation within migratory groups, it is necessary to quantify the relationship between each individual to fully understand group dynamics, social influences, and the resulting overall migration patterns. Despite existing knowledge from theoretical work (*12*–*14*), studying collective movement in wild migratory species has been a major challenge (*15*) because it is nearly impossible to record the simultaneous movements of freely flying animals in large, natural groups with appropriate spatio-temporal resolutions.

Here we approached this question by tagging large numbers of juvenile white storks (*Ciconia ciconia*) with high-resolution tracking devices (Fig. 1A-E). We recorded the trajectories of naturally migrating juvenile storks flying in a flock of 27 GPS-tagged birds (together with untagged birds, Table S1) during the first five days of their migratory journey, over thousand kilometers. Due to birds eventually leaving the flock, flock size was 27, 22, 21, 20 and 17 for the first five days, respectively. Using solar GSM-GPS-ACC loggers, we recorded tri-axial acceleration (10.54Hz for 3.8s every 10min) and high frequency GPS locations (1Hz for 2 or 5min every 15min; synchronized in time between individuals; henceforth: *GPS burst*) of each individual during the group flights (Fig. 1F-I). Following these 5 days, we

"This manuscript has been accepted for publication in Science. This version has not undergone final editing. Please refer to the complete version of record at http://www.sciencemag.org/. The manuscript may not be reproduced or used in any manner that does not fall within the fair use provisions of the Copyright Act without the prior, written permission of AAAS."

continued to monitor each bird's movements throughout their life using GPS and accelerometer recordings at lower resolution to document its entire lifetime (Fig. S1).

Similar to other large-bodied soaring migrants (*16–19*), white storks try to reduce the amount of costly flapping flight by exploiting their atmospheric surroundings (*20*). Here, when comparing movement activity between our tagged juvenile birds of the same flock, we found large differences in their amount of costly flapping. For each bird, we calculated a quantitative measure of animal activity from tri-axial acceleration data (*21*, *22*); henceforth: flapping activity; Fig. 1F; see Methods). Although storks flew in close proximity (Fig. S2-S3), flapping activity ranged from 0.8 to 1.8. Thus, to cover the same distance during the same time, some individuals performed considerably more flapping flight than did others. Flapping activity was not influenced by individual features (e.g. body measures, sex) or conditions prior to fledging (general linear model; $F_{17,8} = 0.798$, $P = 0.671$; Table S2); within-individual differences in flapping activity were stable across the different migration days (Table S3).

First, we examined how these differences in flapping activity relate to birds' positions within the group. Exploring group structure in detail is challenging because of the different flight modes of soaring migrants (*23*, *24*). To examine flock organization during all flight modes, we developed a metric that quantifies time advances/delays $\Delta t$ between each pair of birds allowing us to measure the time that separates two individuals, i.e. how much time a bird needs to reach the current location of the other bird (Fig. S4-S5). Storks with low flapping activity flew on average ahead of other flock members whereas high flapping activity birds flew behind (Pearson's $r = -0.778$, $n = 27$, $P = 1.7 \cdot 10^{-6}$, Fig. 2A). Next, we found that an individual's position within the flock (time advance/delay $\Delta t$) correlated with leadership (*25*) during the gliding segments (Fig. 2B, Pearson's $r = 0.846$, $n = 27$, $P = 2.7 \cdot 10^{-8}$). Further, following birds tended to have a higher flapping activity than did leaders (Pearson's $r = -0.770$, $n = 27$, $P = 2.6 \cdot 10^{-6}$, Fig. S6-S7). Because leader/follower roles are reflected in the front/back positions in the flock, we refer to birds that are on average ahead of the flock as leaders, and those behind as followers (SI text, Fig. S8-S9).

Followers not only spent considerably more time flapping their wings, but they also spent less time thermalling than did leaders ahead (Pearson's $r = -0.688$, $n = 27$, $P = 7.2 \cdot 10^{-5}$, Fig. 2C). Followers finished thermalling earlier, at a lower altitude, likely to avoid being isolated from others, thus seemingly failing to exploit the full potential of thermals (Fig. S10). In addition, followers flew farther behind, and at increasingly lower altitudes than leaders during glides (Fig. 2D,S11-S13). Given that the tagged juveniles migrated together with untagged storks, it is likely that the motion of the observed leaders was in fact affected by other, possibly more experienced, adult birds. Juveniles have higher flight costs than adults, but improve their ability to utilize thermals effectively throughout their journey (*26*). Collective movements may also partly arise from identical reactions to the same environmental features, but here we cannot distinguish between responses to environmental and social cues (*27*).

Leaders and followers differed in their path "tortuosity" while flying within the thermals. Leading birds showed irregular circling while thermalling (calculated as the absolute value of the time derivative of the horizontal curvature, $|d\kappa/dt|$), demonstrating that they exhibit considerable adjustments of their flight paths, consistent with a need to locate the center of the complex thermal structures. By contrast followers circled more regularly, indicating that, as theory suggested (*24*), followers can benefit from social information to reach the center of thermals; time advance/delay correlates highly with $|d\kappa/dt|$ when examining individual averages (Pearson's $r = 0.570$, $n = 27$, $P = 0.002$, Fig. 3A). Because every bird spent some time in the front and back half of the flock, we determined each individual's thermalling performance when ahead and behind the center of the flock (Fig. 3B). Almost all exhibited more regular circling and faster climb rates in thermals when following others than when flying ahead (paired t-test, $n$ = 22, $P$ = 0.030 and $P$ = 0.018, respectively, Fig. 3C,S13-S14).

Examining the complete migratory paths of the 27 birds (at lower temporal resolution) revealed considerable differences in migratory distance, with some birds remaining within Europe and others travelling several thousand kilometers to Africa (Fig 4). Migratory distance was strongly correlated with the birds' migratory flight behavior; those birds that exhibited a high proportion of (costly) flapping activity migrated less far than did those birds that occupied frontal positions, and low flapping activity, when within the flock (Fig 4-inset, Pearson's $r$ = -0.66, $n$ = 20, $P$ = 0.001). These differences in long-term migration behaviors can be predicted using only a few minutes of movement data from the first migration day of the flock (SI text, Fig. S15). Furthermore, flight time prior to migration (i.e. total number of GPS bursts in which each bird was found to be flying, before migrating) is also highly correlated with flapping activity (Pearson's $r$ = -0.648, $n$ = 27, $P$ = $2.6 \cdot 10^{-4}$; Fig. S16) and migratory distance (Pearson's $r$ = 0.619, $n$ = 20, $P$ = 0.004). The differences in flight performance between leaders and followers suggest that juvenile storks may differ in their aerodynamic features and/or their behavioral strategies which, in turn, may affect their migration and group behavior over multiple scales. Nevertheless, birds can compensate for their inferior flight skills (e.g. lower glide ratio, higher ratio of flapping flight) by following others, which enables them to rise faster within thermals (Fig. S13-14).

Contrary to storks that form large groups with spatio-temporally dynamic structures, other species have been suggested to improve social information usage by flying in V-formation (*28*). While the number of studies that use advanced tracking technologies that examine collective migration increases (*3*, *4*, *29*), the consequences of social behavior and social organization are still largely unknown; especially in wild, freely moving animals. Here we identified two different behavioral strategies in a flock of migrating white storks; a finding that is also in agreement theoretical predictions (*2*). We unraveled mechanisms of collective migration in a natural environment by showing how local scale leader/follower strategies emerge through a differential exploitation of the atmosphere. We suggest that integrating intra-specific interactions into the study of animal movements will enable a better, more mechanistic understanding of broad scale ecological processes.

**Acknowledgments:** We thank all people that helped during fieldwork, especially Wolfgang Schäfle, Riek van Noordwijk, Bart Kranstauber, Daniel Piechowski, Babette Eid and Yvonne Flack. We thank Wolfgang Heidrich and Franz Kümmeth (e-obs, Munich, Germany) for their suggestions on logger programming, and Sarah Davidson for setting up the movebank.org data repository.
Funding: We acknowledge funding from the Max Planck Institute for Ornithology. A.F. was supported by the German Aerospace Center (DLR) and the Christiane Nüsslein-Volhard Stiftung; M.N. received additional funding from the Royal Society Newton Alumni scheme.

Author contributions: A.F. and M.W. conceived the idea and designed the project; A.F, W.F and M.W. conducted fieldwork and collected the data; M.N. designed the collective trajectory analyses and visualizations; A.F. and M.N. designed the data analyses and analyzed the data with contributions from I.D.C.; A.F. and M.N. wrote the text with contributions from I.D.C.

Competing interests: The authors declare that they have no competing interests.

Data and materials availability: The data that were used for this study are part of the MPIO white stork lifetime tracking data (Flack et al. 2016) and is available through the Movebank Data Repository (https://www.movebank.org/node/15294 DOI: **10.5441/001/1.bj96m274**)

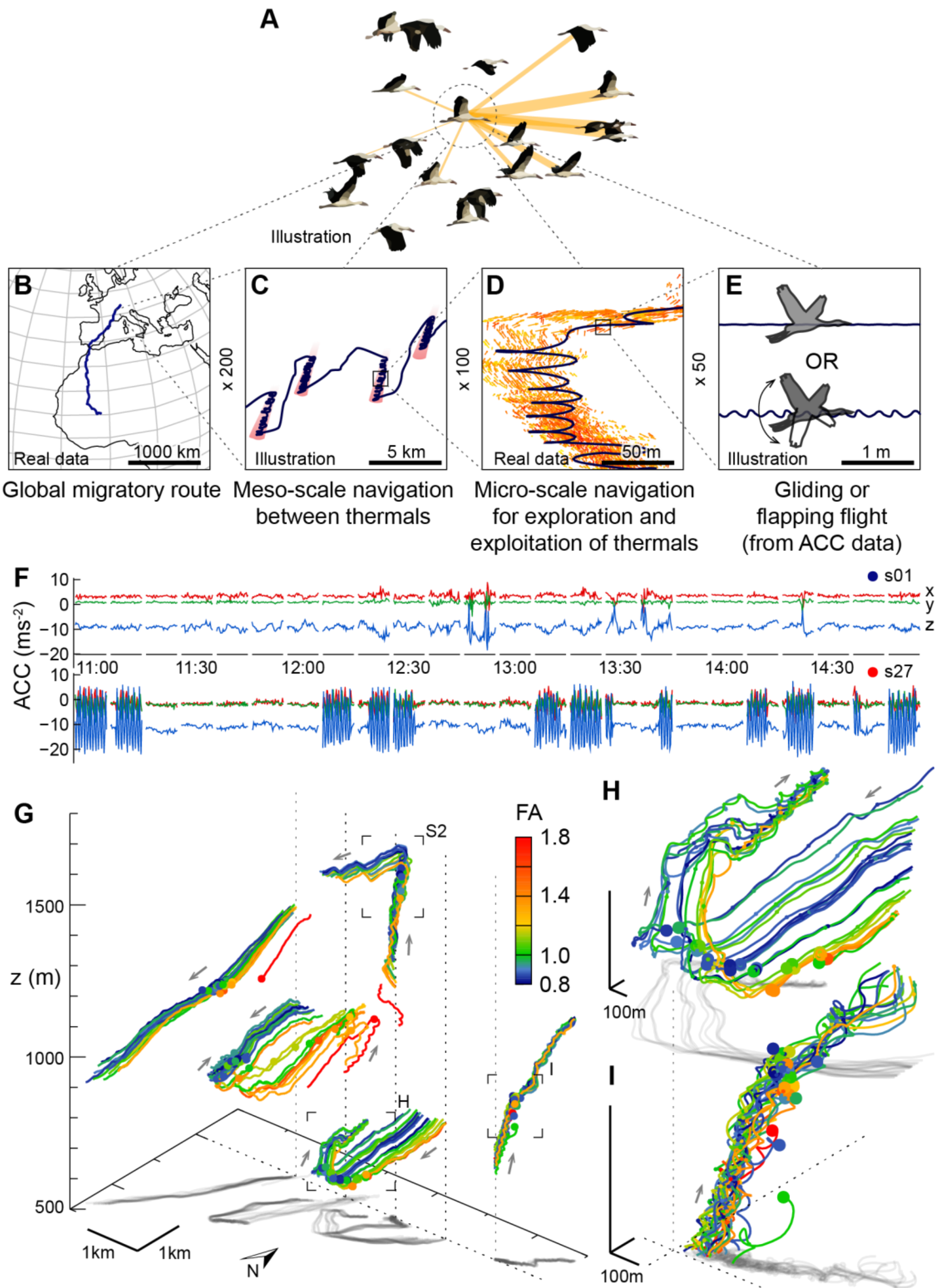

**Fig. 1: Collective migration on different spatial scales (A-E) recorded using acceleration (F) and high-resolution GPS data (G-I). A-E)** Social interactions during migration (A) shape the global migratory route (B) by influencing small scale navigational decisions (i.e. flight behavior between (C) and within (D) thermals), and individual flight performance (E). **F**) Sample of tri-axial accelerometer data used to calculate flapping activity defined as the standardized mean of daily ODBA. Plots show birds with lowest (s01; top) and highest (s27) flapping activity. **G**) Five flock trajectories (1Hz GPS bursts) of migrating storks during thermalling and gliding flight. Bursts are shifted by 1km for visualization. Grey arrows indicate flight direction. Filled circles show the positions of all individuals at 2min. Track color corresponds to flapping activity. **H-I**) Enlarged view of the tracks marked in (G).

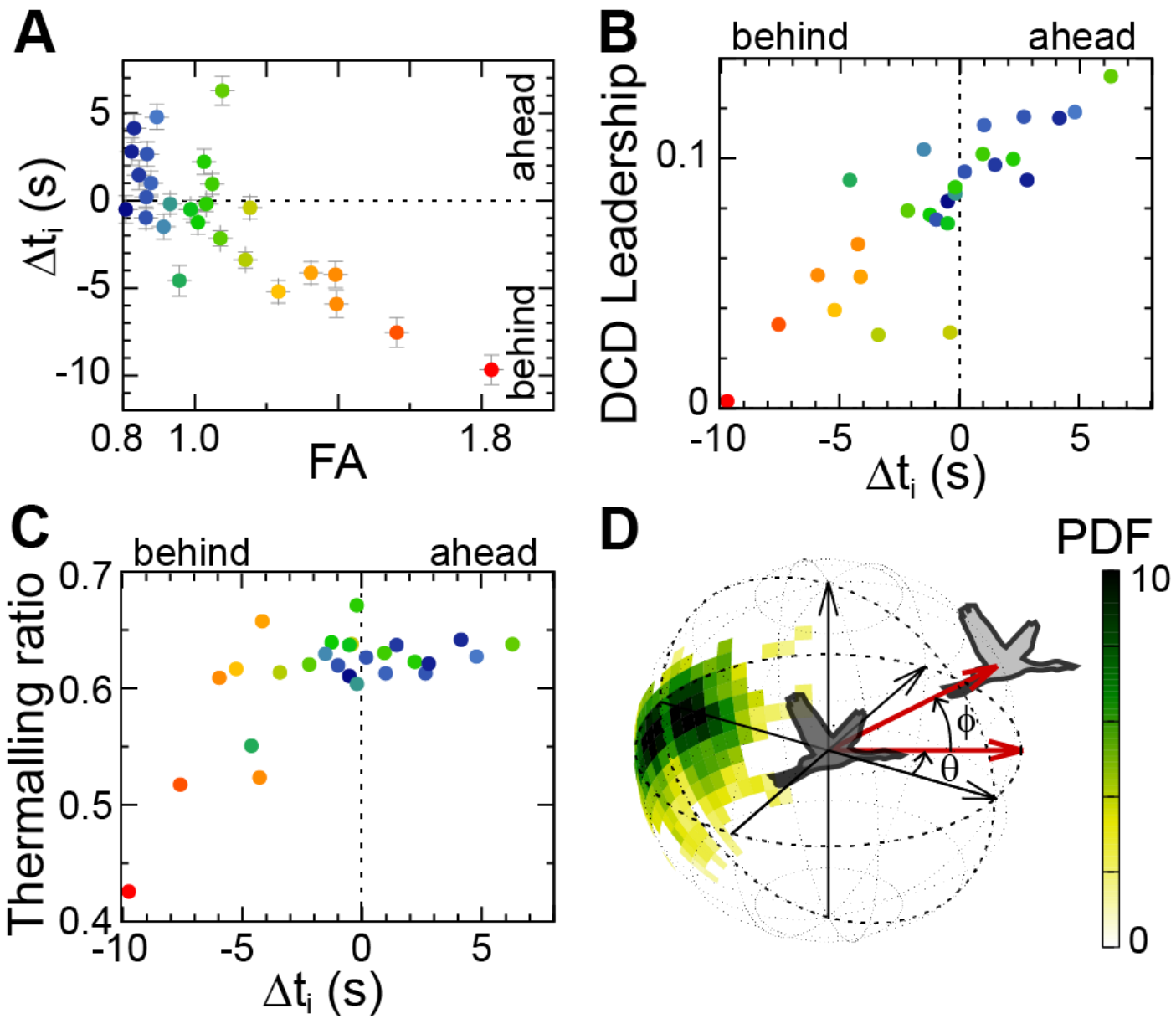

**Fig. 2. Relationship between time advance/delay, leadership, and flapping activity; and the relative positions between leaders and followers.** **A**) Individual time advance/delay ($\Delta t$) averaged over all pairs and bursts against flapping activity. Bars represent standard error of the mean. Color coding shows flapping activity (A-C). **B**) Directional correlation delay leadership (DCD) during gliding against $\Delta t$. **C)** Average thermalling ratio in the first 2min of the bursts against $\Delta t$. **D**) Relative position of followers in a spherical coordinate system during gliding. We placed the focal bird (dark grey - one of five birds with lowest flapping activity) at the center of the coordinate system and measured the relative position of another bird (light grey - one of five birds with highest flapping activity) using the polar ($\vartheta$) and elevation angle ($\phi$) both measured from the focal bird's horizontal flight direction. Probability density function (PDF) shows the relative location between these birds when both were gliding and $\Delta t \in$ [2.5s,7.5s]).

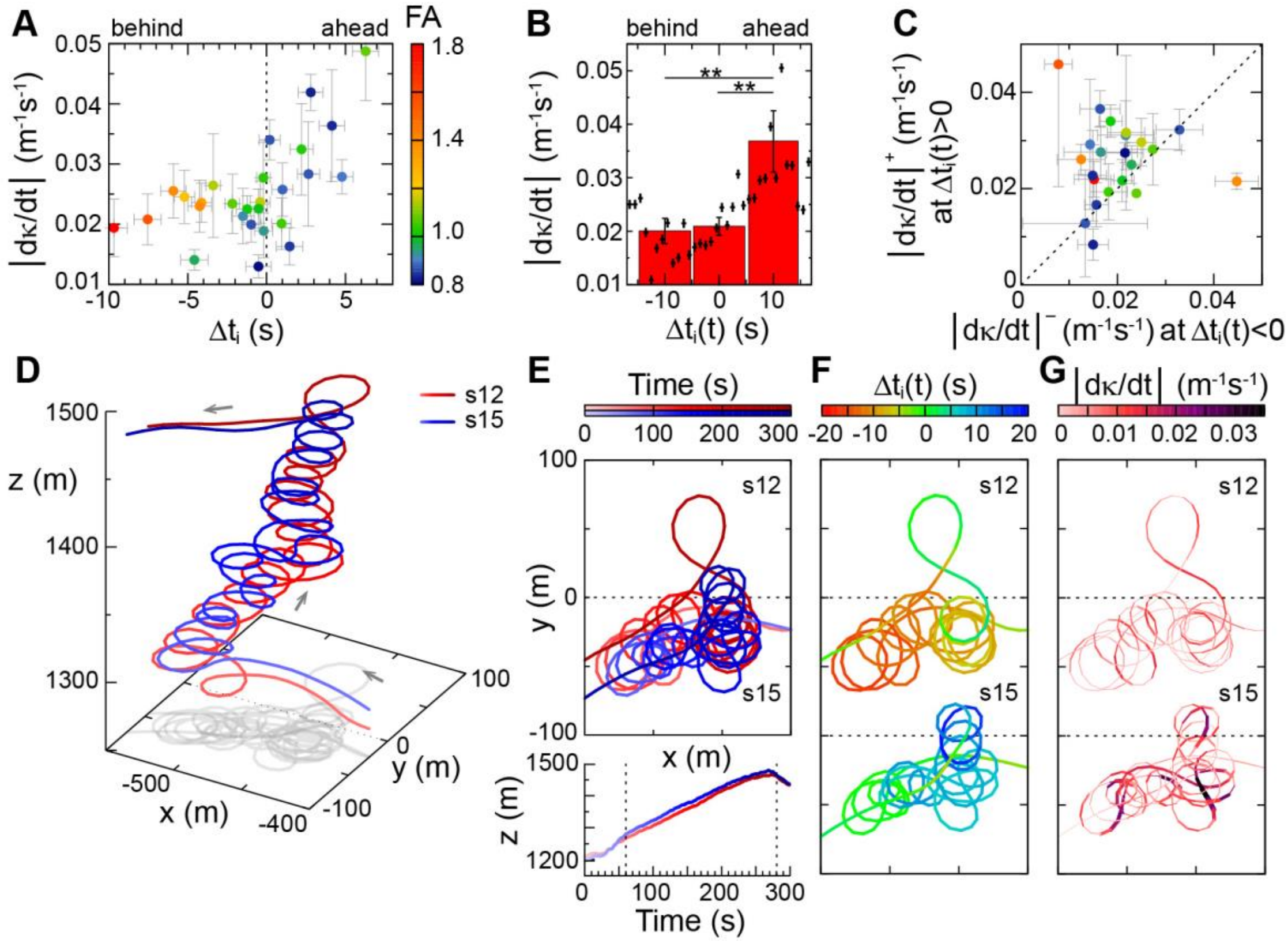

**Fig. 3. Derivative of curvature during thermalling flight for leaders and followers. A**) Relationship between the absolute value of the time derivative of horizontal path curvature ($|d\kappa/dt|$) and time advance/delay $\Delta t$; averaged for each individual across the first migration day. Bars represent standard error (SE) of the mean. **B**) $|d\kappa/dt|$ calculated using $\Delta t$ as 1s (crosses) and 10s (bars) bins. Error bars represent standard deviation of the mean (** $P < 0.01$, $n = 27$, two-tailed t-test). **C**) Relationship between $|d\kappa/dt|$ while flying ahead ($\Delta t > 0$) and behind ($\Delta t < 0$) the average of the flock. Line shows $y = x$. Bars represent standard error (SE) of the mean. **D**) Example trajectories for illustrating the derivative of curvature. Of all bursts that contained thermalling and had a small wind drift, we chose one random example and depicted the most leading (in blue) and following (in red) individuals identified based on their time advance/delay (highest and lowest value of that burst, respectively). **E**) Horizontal (top) and vertical (bottom) movement components of these tracks. **F-G**) Tracks color-coded to show $\Delta t$ (F) and $|d\kappa/dt|$ (G; s15 shifted down for better visualization). $|d\kappa/dt|$ is also indicated by line width.

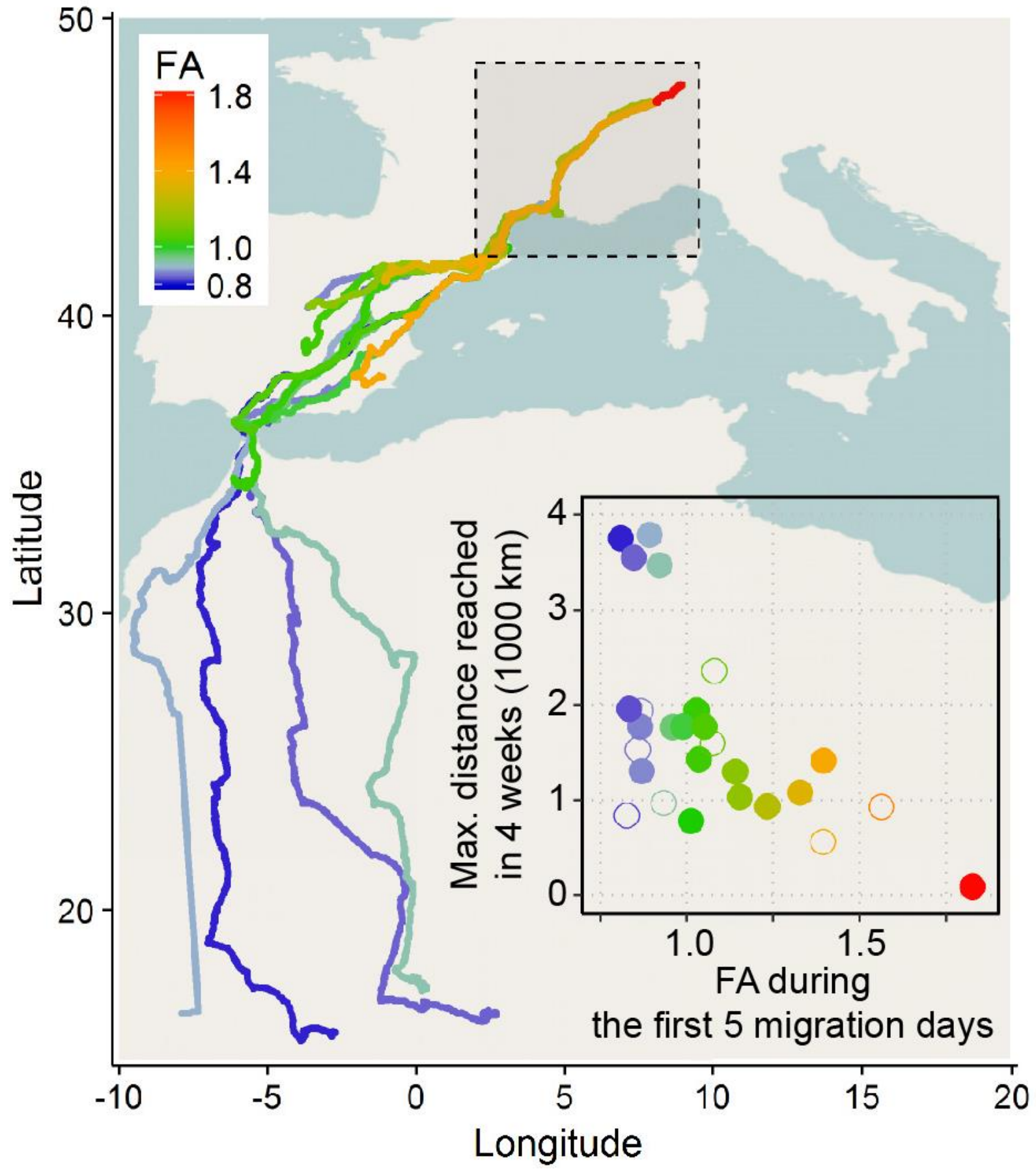

**Fig. 4. Relationship between migratory distance and flapping activity**. **A**) Migration routes of storks during the first 4 weeks of migration. Tracks are color-coded based on the overall flapping activity measured during the high-resolution data acquisition period (highlighted as black rectangle). Bottom right inset shows relationship between flapping activity and maximum distance reached within 4 weeks. Color corresponds to flapping activity; open circles show birds that died within the first 4 weeks.